\documentclass[aps,prd,onecolumn,groupedaddress,showpacs,nofootinbib,amssymb]{revtex4-2}
\usepackage[dvips]{graphicx}
\usepackage{amssymb}
\usepackage{amsmath}
\usepackage{graphicx,,color}
\usepackage{amsfonts}
\usepackage{bm}
\usepackage{cancel}
\usepackage{comment}

\newcommand\be{\begin{equation}}
\newcommand\ee{\end{equation}}

\allowdisplaybreaks[4]

\begin{document}

\tolerance=5000

\title{Non-minimal Derivative Coupling Theories Compatible with GW170817}
\author{V.K. Oikonomou,$^{1,2}$}\email{voikonomou@gapps.auth.gr;v.k.oikonomou1979@gmail.com}
\affiliation{$^{1)}$Department of Physics, Aristotle University of
Thessaloniki, Thessaloniki 54124, Greece \\ $^{2)}$L.N. Gumilyov
Eurasian National University - Astana, 010008, Kazakhstan}

 \tolerance=5000

\begin{abstract}
In this work we aim to revive the interest for non-minimal
derivative coupling theories of gravity, in light of the GW170817
event. These theories include a string motivated non-minimal
kinetic term for the scalar field of the form $\sim \xi(\phi)
G^{\mu\nu}\partial_\mu\phi\partial_\nu\phi$ and predict that the
primordial tensor perturbations have a speed that it is distinct
from the speed of light. Due to the fact that the Universe is
classical during and in the post-inflationary epoch, there is no
fundamental reason for the graviton to change its mass, the
GW170817 event severely constrained these theories. We analyze and
formalize the inflationary phenomenology of these theories, using
both the latest Planck data and the GW170817 event data to
constrain these theories. Due to the fact that there is no
constraint in the choice of the scalar potential and the
non-minimal coupling function, we provide several classes of
viable models, using convenient forms of the minimal coupling
function in terms of the scalar potential, aiming for analyticity,
and we discuss the advantages and disadvantages of each viable
model.
\end{abstract}


\maketitle

\section{Introduction}

Modern cosmology is rendered a precision science, due to the
astonishing amounts of observational data available nowadays.
Thus, theoretical models of the Universe are currently being
severely scrutinized, and the years to follow are promising
regarding cosmological theories. The year that utterly changed the
theoretical perception in theoretical cosmology was 2017 with the
remarkable GW170817 event from the LIGO-Virgo collaboration
\cite{TheLIGOScientific:2017qsa,Monitor:2017mdv,GBM:2017lvd,LIGOScientific:2019vic},
which was astonishing due to a fact that apart from the
gravitational wave detected by the merging neutron stars, a
kilonova was also detected. This event changed our theoretical
perception of our Universe, since massive gravity theories of
inflation were excluded or severely constrained
\cite{Ezquiaga:2017ekz,Baker:2017hug,Creminelli:2017sry,Sakstein:2017xjx}.
The GW170817 event started the chorus of groundbreaking
observations, continued in 2023 with the NANOGrav detection of a
stochastic gravitational wave background \cite{NANOGrav:2023gor}.
These collaborations, in conjunction with the future observations
that are highly anticipated, coming from gravitational wave
experiments like LISA \cite{Baker:2019nia,Smith:2019wny}, BBO
\cite{Crowder:2005nr,Smith:2016jqs} and DECIGO
\cite{Seto:2001qf,Kawamura:2020pcg}, are expected to further
enlarge our knowledge about astrophysical and cosmological
theories. Regarding the GW170817 event, as we mentioned, it
severely constrained theories that predict gravitational waves
with speed smaller than that of light. This is a late-time event
though, thus one might claim that the gravitational wave speed at
early times could be different from that of light. However, the
primordial gravitational waves, are basically classical tensor
perturbations of a classical spacetime, describing a classical
post-Planck Universe. Indeed, when the inflationary era is
believed to have started, the Universe was four dimensional and
free from severe quantum effects. The quantum effects may have
their imprints on the inflationary Lagrangian, however, no direct
quantum effect is expected when the inflationary era started, and
beyond that. Thus, from a particle physics viewpoint, there is no
fundamental inherent mechanism that can effectively change the
graviton mass during and in the post-inflationary era. Thus the
primordial gravitational waves should also be constrained in the
same manner as ordinary late-time gravitational waves from the
GW170817 event. Motivated by this aspect of research, in this work
we shall thoroughly study a class of theories which was affected
by the GW170817 event, the non-minimal derivative theories of
gravity which include non-minimal kinetic coupling terms of the
form $\sim \xi(\phi) G^{\mu\nu}\partial_\mu\phi\partial_\nu\phi$.
These theories constitute an important subclass of Horndeski
theories
\cite{horndeskioriginal,Kobayashi:2019hrl,Kobayashi:2016xpl,Crisostomi:2016tcp,Bellini:2015xja,Gleyzes:2014qga,Lin:2014jga,Deffayet:2013lga,Bettoni:2013diz,Koyama:2013paa,Starobinsky:2016kua,Capozziello:2018gms,BenAchour:2016fzp,Starobinsky:2019xdp},
and were thoroughly studied in the pre-GW170817 epoch
\cite{Hwang:2005hb,Capozziello:1999xt,Capozziello:1999uwa,Sushkov:2009hk,Minamitsuji:2013ura,Saridakis:2010mf,Barreira:2013jma,Sushkov:2012za,Barreira:2012kk,Skugoreva:2013ooa,Gubitosi:2011sg,Matsumoto:2015hua,Deffayet:2010qz,Granda:2010hb,Matsumoto:2017gnx,Gao:2010vr,Granda:2009fh,Germani:2010gm,Fu:2019ttf},
see also \cite{Petronikolou:2021shp} where the same non-minimal
derivative coupling term is suitably chosen in order to have
constant gravitational wave speed and equal to the light speed and
also to solve the $H_0$-tension, and also Ref.
\cite{Karydas:2021wmx} in which case the non-minimal derivative
coupling is used for inflation but in a way that it becomes
negligible after inflation exit, and thus passing GW170817
constraints easily. In this work we aim to revive these theories,
by also including the constraints from the GW170817 event in their
inflationary phenomenological analysis. A similar approach was
performed in Refs.
\cite{Dima:2017pwp,Kreisch:2017uet,Arai:2017hxj} for Horndeski
theories. Non-minimal derivative theories are basically string
inspired theories, which take into account imprints of a more
fundamental theory on the inflationary Lagrangian, and serve as an
important path of the classical theory towards to the more
fundamental quantum theory. Getting to the details of model
building of the non-minimal derivative kinetic coupling theories,
the scalar field potential and the kinetic coupling scalar
function are unconstrained from a theoretical point of view, so in
principle any choice for these functions can be examined regarding
the phenomenological viability of the theory. In this work we aim
to thoroughly formalize these theories and study their
inflationary phenomenology in a concrete and self-consistent way.
We present various classes of solutions and choices for the
non-minimal kinetic coupling function $\xi(\phi)$ and we examine
the inflationary phenomenology of the theory, confronting the
models with the latest Planck data, and also discussing the
attributes and drawbacks of each model.

This paper is organized as follows: In section II we present the
formalism of non-minimal derivative kinetic coupling theories, we
extract analytic forms for the gravitational wave speed of tensor
perturbations, the slow-roll indices, the observational indices
and the amplitude of the scalar perturbations and we also provide
all the necessary theoretical tools and constraints that we will
enable us to extract a viable phenomenology from these theories.
In section III we study several classes of viable theoretical
models and we discuss the advantages and disadvantages of each
model. Finally, the conclusions follow at the end of the article.

\section{Non-minimal Derivative Coupling Theories Theoretical Framework and Compatibility with GW170817 Constraints}

In principle, the non-minimal derivative coupling theories
constitute a string-corrected scalar theory, with leading order
corrections in the Regge slope, without including
Einstein-Gauss-Bonnet terms and higher derivatives of the scalar
field too \cite{Hwang:2005hb,Cartier:2001is}. The gravitational
action reads,
\begin{equation}
\centering \label{action}
\mathcal{S}=\int{d^4x\sqrt{-g}\left(\frac{R}{2\kappa^2}-\frac{1}{2}g^{\mu\nu}\partial_\mu\phi\partial_\nu\phi-V(\phi)-c\xi(\phi)
G^{\mu\nu}\partial_\mu\phi\partial_\nu\phi\right)}\, ,
\end{equation}
with $R$ denoting as usual the Ricci scalar, $g$ stands for the
determinant of the metric tensor, $\kappa=\frac{1}{M_P}$ where
$M_P$ is the reduced Planck mass, $V(\phi)$ is the scalar field
potential, and $\xi(\phi)$ is the non-minimal derivative coupling
function of the kinetic coupling term $\sim c\,\xi(\phi)
G^{\mu\nu}\partial_\mu\phi\partial_\nu\phi$, where $G^{\mu\nu}$
denotes the Einstein tensor
$G^{\mu\nu}=R^{\mu\nu}-\frac{1}{2}g^{\mu\nu}R$, and $c$ is an
arbitrary parameter with mass dimensions $[M]^{-2}$.

With regard to the background metric, we shall consider a flat
Friedman-Lemaitre-Robertson-Walker (FLRW) background with line
element,
\begin{equation}
\centering \label{metric} ds^2=-dt^2+a^2\delta_{ij}dx^idx^j\, ,
\end{equation}
with $a(t)$ denoting the scale factor of the Universe. In this
type of theories, the propagation speed of the tensor
perturbations, or equivalently the speed of the primordial
gravitational waves, has the following form \cite{Hwang:2005hb},
\begin{equation}
\centering \label{cT} c_T^2=1-\frac{Q_f}{2Q_t}\, ,
\end{equation}
with the term $Q_f$ being equal to,
\begin{equation}\label{extraqf}
Q_f=4c\xi\dot\phi^2\, ,
\end{equation}
where the ``dot'' denotes differentiation with respect to the
cosmic time hereafter. Moreover, $Q_t$ in Eq. (\ref{extraqf})
stands for $Q_t=\frac{1}{\kappa^2}+c\xi\dot\phi^2$. Thus the
primordial gravitational wave speed takes the form,
\begin{equation}
\centering \label{cTexplicit}
c_T^2=1-\frac{2c\kappa^2\xi\dot\phi^2}{1+c\kappa^2\xi\dot\phi^2}\,
.
\end{equation}
Thus in order to comply with the striking GW170817 event which
rules out massive graviton theories, it is vital that the
following constraint is satisfied at first horizon crossing during
inflation, but also for all the subsequent eras,
\begin{equation}\label{mainconstraint}
\kappa^2c\xi\dot{\phi}^2\ll 1\, ,
\end{equation}
and specifically the term $\kappa^2c\xi\dot{\phi}^2$ must be
sufficiently smaller than unity in order for the GW170817
experimental constraint to be satisfied, which is,
\begin{align}
\label{GWp9} \left| c_T^2 - 1 \right| < 6 \times 10^{-15}\, ,
\end{align}
in natural units. Regarding the constraint (\ref{mainconstraint}),
as we mentioned, it has to be satisfied for all the post-first
horizon crossing eras, and this is a necessity since there is no
fundamental particle physics reason for the graviton to change its
mass in the inflationary and post inflationary era, because the
inflationary era and all the subsequent eras, are basically
classical eras of our Universe, hence no big change in the
propagating mass of the particles is expected. The gravitational
field equations can be obtained by varying the gravitational
action (\ref{action}), with respect to the metric tensor and with
respect to the scalar field, so we get the Friedmann equation,
\begin{equation}
\centering \label{motion1}
\frac{3H^2}{\kappa^2}=\frac{1}{2}\dot\phi^2-9c\, \xi
H^2\dot\phi^2\, ,
\end{equation}
the Raychaudhuri equation,
\begin{equation}
\centering \label{motion2} -\frac{2\dot
H}{\kappa^2}=\dot\phi^2-8H^2(\ddot\xi-H\dot\xi)+c\left(2\xi(\dot
H-3H^2)\dot\phi^2+4H\xi\dot\phi\ddot\phi+2H\dot\xi\dot\phi^2\right)\,
,
\end{equation}
and the scalar field generalized Klein-Gordon equation,
\begin{equation}
\centering \label{motion3}
\ddot\phi+3H\dot\phi+V'+6c\left(H^2(\dot\xi\dot
H+2\xi\ddot\phi)+2H(2\dot H+3H^2)\xi\dot\phi\right)=0\, .
\end{equation}
The field equations are coupled differential equations with quite
involved form, and it is understandable that an analytic solution
is not easy to be obtained. However, several simplifications may
be applied during the inflationary era, such as the slow-roll
approximation for the scalar field and the inflationary conditions
for the Hubble rate, and specifically the following
approximations,
\begin{align}
\centering \label{slowroll} \dot H&\ll H^2 &\frac{1}{2}
\dot\phi^2&\ll V& \ddot\phi&\ll H\dot\phi\, .
\end{align}
Also we must take into account the condition of Eq.
(\ref{mainconstraint}), which is a imposed inherent necessity for
the viability and compatibility of the theory with the GW170817
event, thus we must also take into account that $\kappa^2c\,
\xi\dot{\phi}^2\ll 1$. This term, namely $9c\, \xi H^2\dot\phi^2$
enters the Friedmann equation and thus it is not dominant at
leading order, and also $\dot{\phi}^2$ must also be subdominant,
hence only the scalar potential survives, therefore the Friedmann
equation becomes,
\begin{equation}
\centering \label{motion4} H^2\simeq \frac{\kappa^2V}{3}\, .
\end{equation}
In the end, the approximations we made, must be checked
explicitly, namely,
\begin{equation}\label{approximationsnatural}
c\, \xi\dot{\phi}^2\ll V,\,\,\,\dot{\phi}^2\ll V\, ,
\end{equation}
however these two are not additional constraints in the theory,
imposed by hand, but these stem from the natural requirements that
the gravitational wave speed is equal to that of light and also
that the slow-roll approximation holds for the scalar field. For
consistency though, we will validate these constraints for any
viable model we shall present. Accordingly, by using the slow-roll
and inflationary constraints, the Raychaudhuri equations read,
\begin{equation}
\centering \label{motion5}
-\dot{H}=\frac{\kappa^2\dot{\phi}^2}{2}-3c\kappa^2\xi
H^2\dot{\phi}^2\, ,
\end{equation}
or equivalently,
\begin{equation}
\centering \label{Raychaudhuri}
-\dot{H}=\frac{\kappa^2\dot{\phi}^2}{2}\left(1-6c\, \xi H^2
\right)\, .
\end{equation}
Accordingly, by using the slow-roll and inflationary constraints
(\ref{slowroll}), the modified Klein-Gordon equation reads,
\begin{equation}
\centering \label{motion6} V'+3H\dot{\phi}+ 36 c \xi
\dot{\phi}H^3\simeq 0\, ,
\end{equation}
which can be rewritten,
\begin{equation}\label{kleingordon}
\dot{\phi}=-\frac{V'}{3H\left(1+12c\, \xi H^2\right)}\, .
\end{equation}
From the final form of the Raychaudhuri equation
(\ref{Raychaudhuri}) and of the modified Klein-Gordon equation
(\ref{kleingordon}), it is obvious that the term $\sim c\, \xi
H^2$ enters both equations and if this dominant compared to unity,
then both the Raychaudhuri equation and the modified Klein-Gordon
equation can be simplified. We shall make this approximation in
the following and also we shall consider the generalized formulas.
So in the following we shall consider two classes of non-minimal
derivative couplings models, one that the following constraint
holds true,
\begin{equation}\label{constraintclass}
c\, \xi H^2\gg 1\, ,
\end{equation}
in which case, the Raychaudhuri and the modified Klein-Gordon
equations take the form,
\begin{equation}
\centering \label{Raychaudhuriconstrained} -\dot{H}=-3c\, \xi
H^2\kappa^2\dot{\phi}^2\, ,
\end{equation}
\begin{equation}\label{kleingordonconstrained}
\dot{\phi}=-\frac{V'}{36c\, \xi H^3}\, ,
\end{equation}
and general models which do not respect the constraint
(\ref{constraintclass}) and thus the Raychaudhuri and the
Klein-Gordon equations are given in Eqs. (\ref{Raychaudhuri}) and
(\ref{kleingordon}) respectively. We shall refer to the models
that satisfy the constraint (\ref{constraintclass}) as
``constrained models''. Now let us present the slow-roll indices
of inflation, the observational indices of inflation and the
relation that yields the $e$-foldings number for the non-minimal
derivative coupling theories of gravity, for both the constrained
models and the unconstrained models. The slow-roll indices for the
non-minimal derivative coupling theories have the general form
\cite{Hwang:2005hb},
\begin{align}
\centering \label{indices} \epsilon_1&=-\frac{\dot
H}{H^2}&\epsilon_2&=\frac{\ddot\phi}{H\dot\phi}&\epsilon_3&=0&\epsilon_4&=\frac{\dot
E}{2HE}&\epsilon_5&=\frac{Q_a}{2HQ_t}&\epsilon_6&=\frac{\dot
Q_t}{2HQ_t}\, ,
\end{align}
where $Q_a=4c\, \xi H\dot\phi^2$, $Q_b=2c\, \xi \dot{\phi}^2$,
$E=\frac{1}{(\kappa\dot\phi)^2}\left(
\dot\phi^2+\frac{3Q_a^2}{2Q_t}+Q_c\right)$, $Q_c=-6c\,
\xi\dot\phi^2H^2$, $Q_d=-4c\, \xi \dot{\phi}^2\dot{H}$,
$Q_e=4c\dot{\phi}\left(\dot{\xi}\dot{\phi}+2\xi\ddot{\phi}-2\xi\dot{\phi}H
\right)$, $Q_f=4c\, \xi \dot{\phi}^2$ and recall that
$Q_t=\frac{1}{\kappa^2}+\frac{Q_b}{2}$. The observational indices
of inflation in terms of the slow-roll indices have the following
form,
\begin{equation}\label{spectralindex}
n_{\mathcal{S}}=1+\frac{2 (-2
\epsilon_1-\epsilon_2-\epsilon_4)}{1-\epsilon_1}\, ,
\end{equation}
regarding the spectral index of the scalar primordial curvature
perturbations, while the tensor-to-scalar ratio has the form,
\begin{equation}\label{tensortoscalar}
r=\left |\frac{16 \left(c_A^3 \left(\epsilon_1-\frac{1}{4} \kappa
^2 \left(\frac{2
Q_c+Q_d}{H^2}-\frac{Q_e}{H}+Q_f\right)\right)\right)}{c_T^3
\left(\frac{\kappa ^2 Q_b}{2}+1\right)}\right |\, ,
\end{equation}
where $c_A$ is the sound speed of the scalar perturbations,
defined as,
\begin{equation}\label{soundspeed}
c_A=\sqrt{\frac{\frac{Q_a Q_e}{\frac{2}{\kappa ^2}+Q_b}+Q_f
\left(\frac{Q_a}{\frac{2}{\kappa
^2}+Q_b}\right)^2+Q_d}{\dot{\phi}^2+\frac{3 Q_a^2}{\frac{2}{\kappa
^2}+Q_b}+Q_c}+1}\, ,
\end{equation}
and also $c_T$ is the speed of tensor perturbations defined in Eq.
(\ref{cTexplicit}). Also the tensor spectral index is given in
terms of the slow-roll parameters as follows,
\begin{equation}\label{tensorspectralindex}
n_{\mathcal{T}}=-2\frac{\left( \epsilon_1+\epsilon_6
\right)}{1-\epsilon_1}\, .
\end{equation}
Let us now express the $e$-foldings number in terms of the
potential and its derivatives. The definition of the $e$-foldings
number is,
\begin{equation}
\label{efolds} \centering
N=\int_{t_i}^{t_f}{Hdt}=\int_{\phi_i}^{\phi_f}\frac{H}{\dot{\phi}}d\phi\,
,
\end{equation}
where $\phi_i$ and $\phi_f$ are the scalar field values at the
beginning (first horizon crossing) and the end of the inflationary
era. The value of the scalar field at the end of the inflationary
era can be obtained by solving $|\epsilon_1|\sim \mathcal{O}(1)$,
while by solving Eq. (\ref{efolds}) with respect to $\phi_i$ one
may obtain the value of the scalar field at first horizon
crossing, namely $\phi_i$. By using Eqs. (\ref{motion4}) and
(\ref{kleingordon}) we get,
\begin{equation}
\label{efoldsuncostrained} \centering
N=\int_{t_i}^{t_f}{Hdt}=\int_{\phi_f}^{\phi_i}\frac{\kappa^2V\left(
1+4c\kappa^2V\xi\right)}{V'}d\phi\, ,
\end{equation}
for the case of the unconstrained, while in the case of the
constrained models which satisfy the constraint
(\ref{constraintclass}), the $e$-foldings number is given by the
following relation,
\begin{equation}
\label{efoldscostrained} \centering
N=\int_{t_i}^{t_f}{Hdt}=4c\kappa^4\int_{\phi_f}^{\phi_i}\frac{\xi
V^2}{V'}d\phi\, .
\end{equation}
Before closing this section, let us also mention another important
constraint which needs to be taken into account in order to
provide a viable phenomenology, related with the amplitude of the
scalar perturbations $\mathcal{P}_{\zeta}(k_*)$, defined as,
\begin{equation}\label{definitionofscalaramplitude}
\mathcal{P}_{\zeta}(k_*)=\frac{k_*^3}{2\pi^2}P_{\zeta}(k_*)
\end{equation}
which must be evaluated at first horizon crossing during the
inflationary era, denoted as $k_*$, known as the pivot scale
$k_*=0.05$Mpc$^{-1}$, and is highly relevant to the CMB
observations. The latest Planck data \cite{Planck:2018jri}
constrain the amplitude of the scalar perturbations to be
$\mathcal{P}_{\zeta}(k_*)=2.196^{+0.051}_{-0.06}\times 10^{-9}$,
when this is evaluated at the CMB pivot scale. The scalar
amplitude for the scalar perturbations $\mathcal{P}_{\zeta}(k)$ in
terms of the two point function for the curvature perturbations
$\zeta (k)$ through $P_{\zeta}(k)$ which appears in Eq.
(\ref{definitionofscalaramplitude}) is,
\begin{equation}\label{twopointfunctionforzeta}
\langle\zeta(k)\zeta (k')\rangle=(2\pi)^3 \delta^3 (k-k')
P_{\zeta}(k)\, .
\end{equation}
For the non-minimal derivative coupling theories, the amplitude
$\mathcal{P}_{\zeta}(k_*)$ in the slow-roll approximation is
\cite{Hwang:2005hb},
\begin{equation}\label{powerspectrumscalaramplitude}
\mathcal{P}_{\zeta}(k)=\left(\frac{k \left((-2
\epsilon_1-\epsilon_2-\epsilon_4) \left(0.57\, +\log \left(\left|k
\eta \right| \right)-2+\log (2)\right)-\epsilon_1+1\right)}{(2 \pi
) \left(z c_A^{\frac{4-n_{\mathcal{S}}}{2}}\right)}\right)^2\, ,
\end{equation}
evaluated at first horizon crossing, with $z=\frac{a \dot{\phi}
\sqrt{\frac{E(\phi )}{\frac{1}{\kappa ^2}}}}{H (\epsilon_5+1)}$ at
first horizon crossing, and in addition the conformal time at
first horizon crossing is
$\eta=-\frac{1}{aH}\frac{1}{-\epsilon_1+1}$ \cite{Hwang:2005hb}.
In the following we shall also take this into account in order to
generate viable phenomenologies of non-minimal derivative coupling
theories.

\section{Phenomenology of Various Classes of Viable Models}

In this section we shall thoroughly analyze the inflationary
phenomenology of several classes of non-minimal derivative
coupling theories, confronting the models with the Planck
constraints. In principle, for the theories at hand, the scalar
potential $V(\phi)$ and the scalar coupling function $\xi(\phi)$
are free to choose, since these are not fundamentally constrained,
as in the case of constrained Einstein-Gauss-Bonnet gravity
developed in Refs.
\cite{Oikonomou:2021kql,Oikonomou:2020sij,Odintsov:2020sqy}. Thus,
we shall choose them freely, using convenient forms for their
functional form which may lead to analytical results. We found
several classes of models that may provide analytic results, and
in many cases viable phenomenologies. We shall use the Planck
units system in which $$\kappa^2=\frac{1}{8\pi
G}=\frac{1}{M_P^2}=1$$ for convenience. Also for each viable model
presented, we shall verify explicitly that the slow-roll and any
additional constraints that render the gravitational wave speed
equal to that of light are satisfied. We start our analysis with
the constrained models that satisfy the constraint of Eq.
(\ref{constraintclass}) and we continue with the unconstrained
models.

\subsection{Phenomenology of Constrained Models}

\begin{figure}
\centering
\includegraphics[width=25pc]{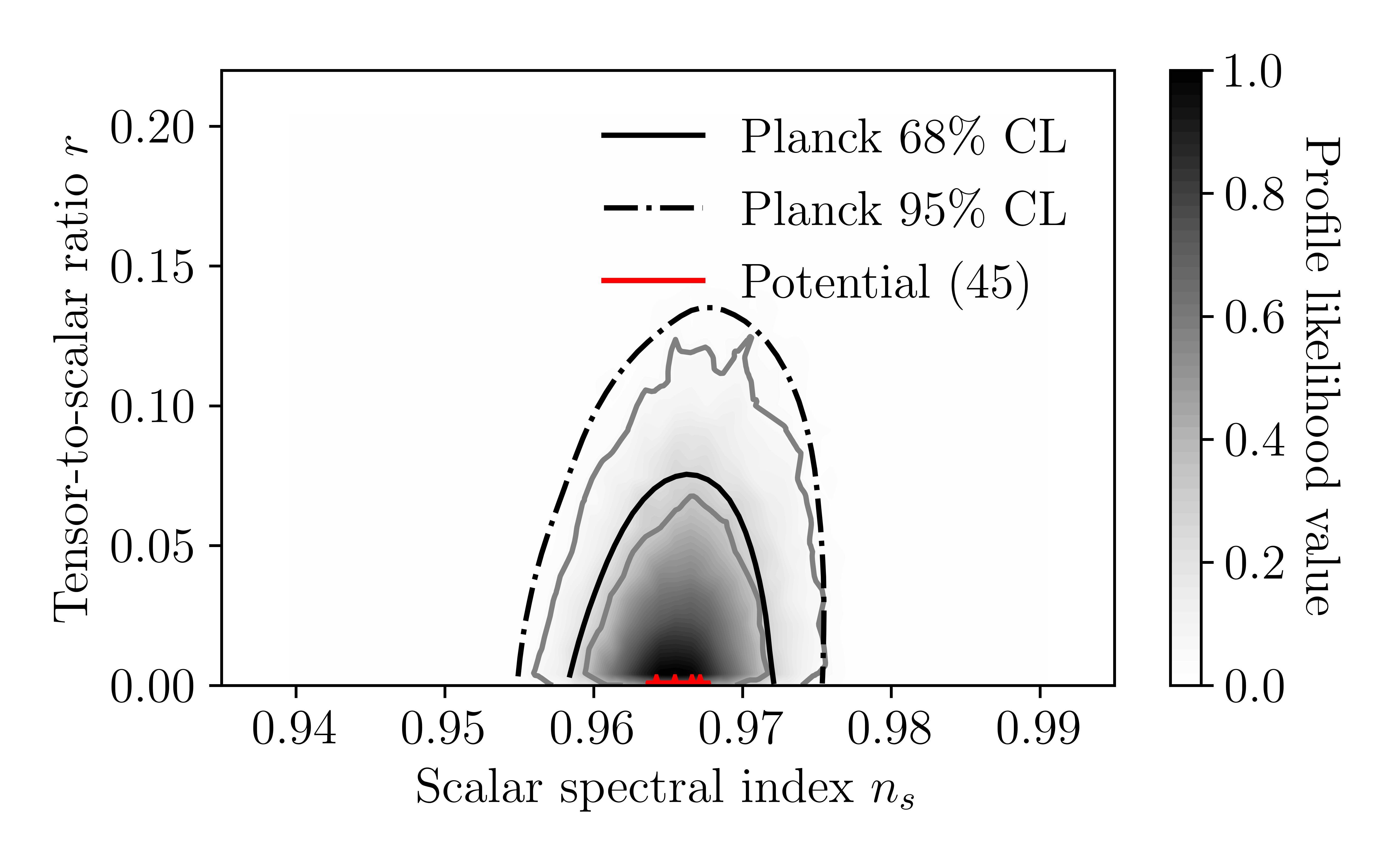}
\caption{Marginalized curves of the Planck 2018 data and the model
with potential $V(\phi)=\frac{M}{1-\frac{d}{\kappa  \phi }}$ of
Eq. (\ref{potential}) vs the Planck 2018 data (red curve),  for
$N=[50,65]$ and for
$(c,d,M,\lambda)=(10^{25.5},10^{-21},1.46725\times
10^{-22},10^{-16.47})$.}\label{plot1}
\end{figure}

\begin{figure}
\centering
\includegraphics[width=25pc]{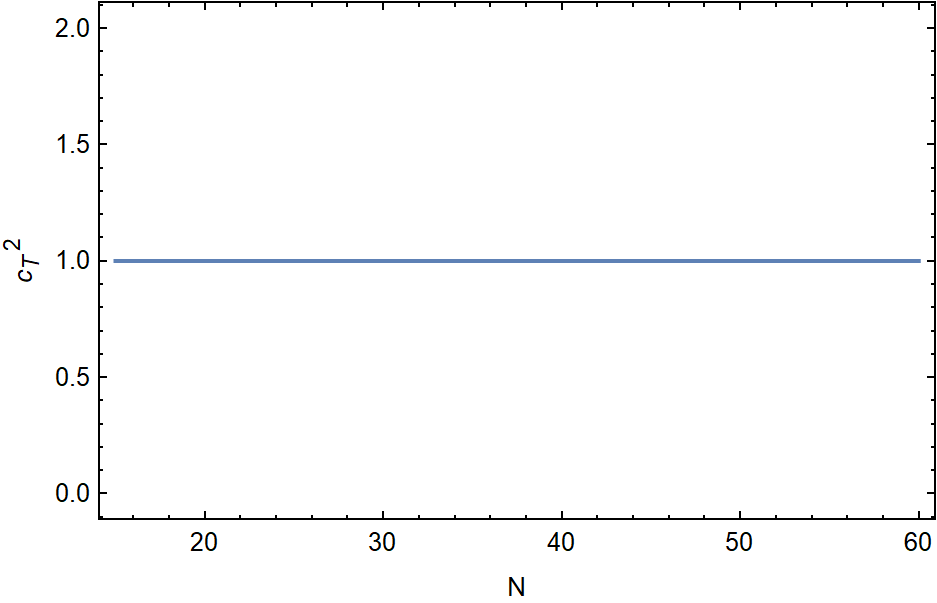}
\caption{The gravitation wave speed $c_T^2$ for the model with
potential $V(\phi)=\frac{M}{1-\frac{d}{\kappa  \phi }}$ of Eq.
(\ref{potential}) for $N=[0,60]$ and for
$(c,d,M,\lambda)=(10^{25.5},10^{-21},1.46725\times
10^{-22},10^{-16.47})$.}\label{plotreff}
\end{figure}

\begin{figure}
\centering
\includegraphics[width=25pc]{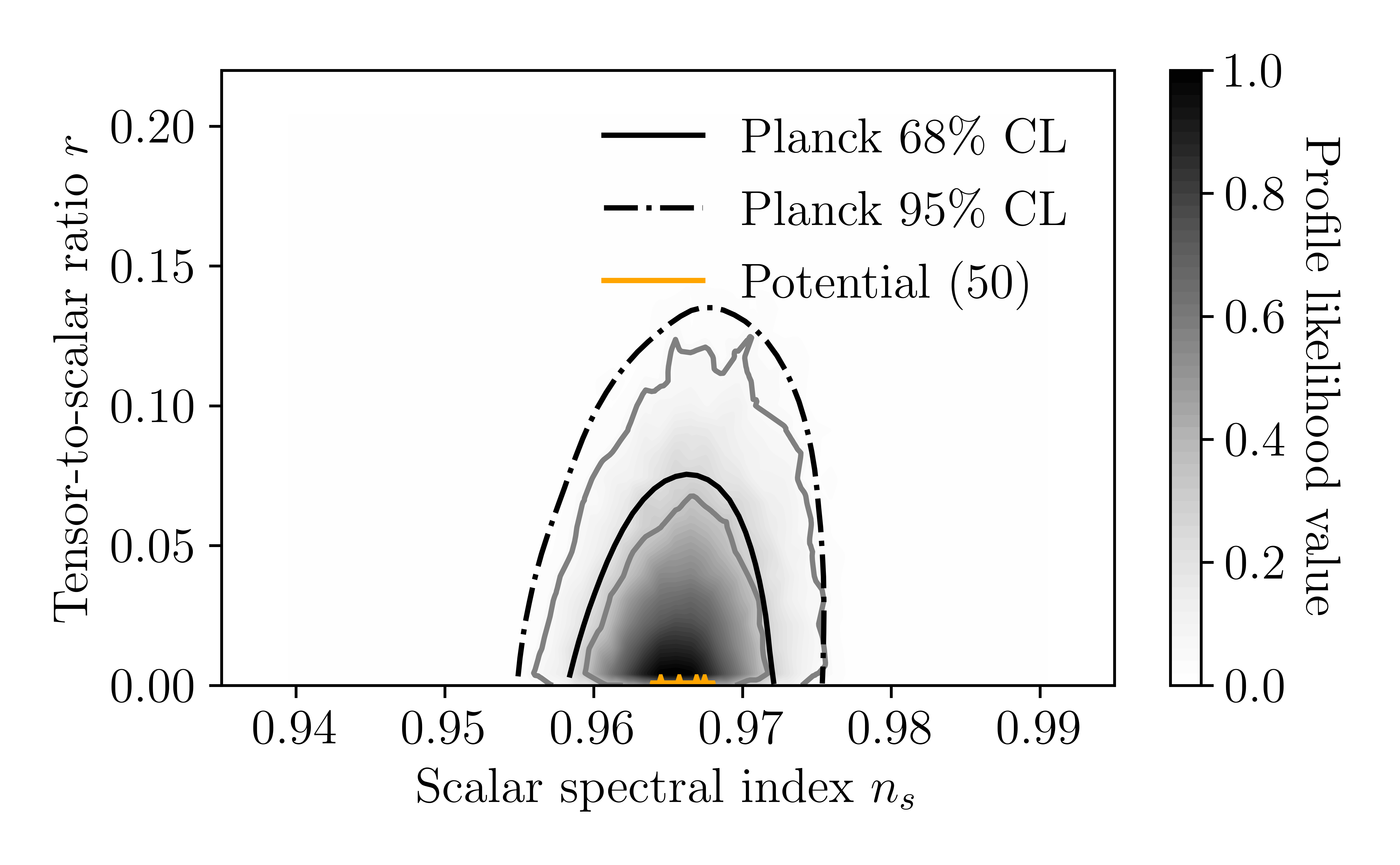}
\caption{Marginalized curves of the Planck 2018 data and the model
with potential $V(\phi)=M\left(1-\frac{d}{\kappa
\phi}\right)^{1/100}$ of Eq. (\ref{potential1}) vs the Planck 2018
data (yellow curve) for $N=[50,65]$ and
$(c,d,M,\lambda)=(10^{25.5},-10^{-21},1.48126\times
10^{-24},10^{-16.47})$.}\label{plot2}
\end{figure}

Let us begin our analysis by studying the constrained models which
must obey the constraint of Eq. (\ref{constraintclass}). For these
models, the Friedmann and Raychaudhuri equations are given in Eqs.
(\ref{motion4}) and (\ref{Raychaudhuriconstrained}) and the
derivative of the scalar field is given by Eq.
(\ref{kleingordonconstrained}), while the $e$-foldings number is
given by Eq. (\ref{efoldscostrained}). Using these, in addition to
Eqs. (\ref{indices}), (\ref{spectralindex}),
(\ref{tensortoscalar}), (\ref{soundspeed}) and
(\ref{tensorspectralindex}), one may obtain study in detail the
inflationary phenomenology of this class of models. As we
mentioned earlier, there is no fundamental constraint that
constrains the functional form of the scalar potential and of the
scalar coupling function, thus these are free to choose and
unrelated. Thus, we shall choose these in a way so that analytic
expressions for the inflationary indices and for the rest of the
parameters are obtained. In order to discover interesting classes
of models, we shall quote here the analytic form of the first
slow-roll index $\epsilon_1$, which is,
\begin{equation}\label{epsilon1analytic}
\epsilon_1=-\frac{V'(\phi )^2}{16 c \kappa ^4 \xi (\phi ) V(\phi
)^3}\, ,
\end{equation}
hence we quote here convenient choices for the coupling function
$\xi(\phi)$ in Planck units:
\begin{equation}\label{couplingfunctionchoices1}
\xi(\phi)=\frac{1}{V(\phi)}\, ,
\end{equation}
\begin{equation}\label{couplingfunctionchoices2}
\xi(\phi)=\lambda  V'(\phi )^2\, ,
\end{equation}
\begin{equation}\label{couplingfunctionchoices3}
\xi(\phi)=\frac{\lambda  V'(\phi )}{V(\phi )^2}\, ,
\end{equation}
\begin{equation}\label{couplingfunctionchoices4}
\xi(\phi)=\frac{\lambda  V'(\phi )}{V(\phi )}\, ,
\end{equation}
\begin{equation}\label{couplingfunctionchoices5}
\xi(\phi)=\frac{\lambda  V'(\phi )}{V(\phi )^3}\, ,
\end{equation}
From all the choices, the most interesting phenomenologically is
the class of models (\ref{couplingfunctionchoices3}). The classes
of models (\ref{couplingfunctionchoices2}),
(\ref{couplingfunctionchoices4}) and
(\ref{couplingfunctionchoices5}) do not yield viable results for a
large number of freely chosen scalar potentials, and the main
reason for this behavior is the final form of the gravitational
wave speed, which can never be compatible with the GW170817
constraint of Eq. (\ref{GWp9}). Indeed for the models of the class
(\ref{couplingfunctionchoices2}), the gravitational wave speed has
the form,
\begin{equation}\label{gwspeedclass2}
c_T^2=\frac{48 c  \lambda  V(\phi )^3-1}{48 c \lambda V(\phi
)^3+1}\, ,
\end{equation}
and as it proves it is always of the order $\sim
\mathcal{O}(10^{-3})$ for a large number of scalar potentials.
Also regarding the class of models
(\ref{couplingfunctionchoices4}) the gravitational wave speed has
the form,
\begin{equation}\label{gwspeedclass2}
c_T^2=\frac{48 c \lambda  V(\phi )^2-V'(\phi )}{48 c \lambda
V(\phi )^2+V'(\phi )}\, ,
\end{equation}
and as it proves in this case too, it is always of the order $\sim
\mathcal{O}(10^{-3})$ for a large number of scalar potentials.
Finally for the class of models (\ref{couplingfunctionchoices5})
the gravitational wave speed has the form,
\begin{equation}\label{gwspeedclass2}
c_T^2=\frac{48 c \lambda -V'(\phi )}{48 c \lambda +V'(\phi )}\, ,
\end{equation}
and as it proves in this case too, it is always of the order $\sim
\mathcal{O}(10^{-3})$ for a large number of scalar potentials.
Hence, we shall analyze only the class of models
(\ref{couplingfunctionchoices1}) and
(\ref{couplingfunctionchoices3}). Among the two, the most
interesting phenomenologically is the class
(\ref{couplingfunctionchoices3}) since we found a large number of
scalar potentials that yield a viable phenomenology, so we start
our analysis with this class of models. In this case, the first
slow-roll index acquires a quite simplified form and it is equal
to,
\begin{equation}\label{epsilon1formclass}
\epsilon_1=-\frac{V'(\phi )}{16 c \lambda  V(\phi )}\, ,
\end{equation}
while the $e$-foldings number as a function of the final and
initial scalar field values reads,
\begin{equation}\label{finalinitialefoldings}
N=4 c \lambda  \phi_f-4 c \lambda  \phi_i\, ,
\end{equation}
and it is independent from the actual form of the potential, due
to the choice of the scalar coupling function
(\ref{couplingfunctionchoices2}). The gravitational wave speed in
this case has the form,
\begin{equation}\label{gwspeedformclass}
c_T^2=\frac{48 c \lambda  V(\phi )-V'(\phi )}{48 c \lambda V(\phi
)+V'(\phi )}\, .
\end{equation}
Among the rest of the slow-roll indices of inflation and the
observational indices, only the tensor spectral index has a simple
form, which we quote here due to the importance of the tensor
spectral index for the primordial gravitational waves,
\begin{equation}\label{tensorspectralindexclass}
n_{\mathcal{T}}=\frac{4 V(\phi ) \left(24 c \lambda V'(\phi
)+V''(\phi )\right)-2 V'(\phi )^2}{\left(16 c \lambda V(\phi
)+V'(\phi )\right) \left(48 c \lambda V(\phi )+V'(\phi )\right)}\,
.
\end{equation}
Now let us consider various scalar potentials and we study their
inflationary phenomenology and consistency in detail. We start
with the following potential,
\begin{equation}\label{potential}
V(\phi)=\frac{M}{1-\frac{d}{ \phi }}\, ,
\end{equation}
in which case the scalar coupling function $\xi(\phi)$ reads,
\begin{equation}\label{xiphi}
\xi(\phi)=-\frac{d \lambda }{M \phi ^2}\, ,
\end{equation}
so the first slow-roll index reads in this case,
\begin{equation}\label{slowrollindexena}
\epsilon_1=-\frac{d}{16 c\, d \lambda  \phi -16 c \lambda \phi
^2}\, ,
\end{equation}
and thus by solving the equation $|\epsilon_1(\phi_f)|=1$ we
obtain the value of the scalar field at the end of inflation
$\phi_f$ which is,
\begin{equation}\label{phif}
\phi_f=\frac{\sqrt{4 c^2 d^2 \lambda ^2+c\, d \lambda }+2 c\, d
\lambda }{4 c \lambda }\, ,
\end{equation}
and from Eq. (\ref{finalinitialefoldings}), the value of the
scalar field at first horizon crossing is,
\begin{equation}\label{phii}
\phi_i=\frac{2 c\, d \lambda +\sqrt{c\, d \lambda  (4 c\, d
\lambda +1)}+N}{4 c \lambda }\, .
\end{equation}
Now a viable phenomenology is obtained for $N=61$ for various
values of the free parameters, for example if we choose the free
parameters of the model as follows
$(c,d,M,\lambda)=(10^{25.5},10^{-21},1.46725\times
10^{-22},10^{-16.47})$, in which case, the spectral index of the
scalar perturbations, the tensor-to-scalar ratio and the tensor
spectral index take the values $n_{\mathcal{S}}=0.967162$,
$r=4.57199 \times 10^{-15}$ and $n_{\mathcal{T}}=-5.74661\times
10^{-16}$, thus the model is viable regarding the observational
indices of inflation. Also in Fig. \ref{plot1} we confront the
model with the Planck likelihood curves, for $N=[50,65]$ and we
can see that the model can be well fitted in the Planck data.
Also the amplitude of the scalar perturbations for the model at
hand for $(c,d,M,\lambda)=(10^{25.5},10^{-21},1.46725\times
10^{-22},10^{-16.47})$ is $\mathcal{P}_{\zeta}(k_*)=2.196\times
10^{-9}$, and actually, the value of the parameter $M$ crucially
affects only the amplitude of the scalar perturbations, since all
the rest of the inflationary indices are independent of $M$. Now,
for $(c,d,M,\lambda)=(10^{25.5},10^{-21},1.46725\times
10^{-22},10^{-16.47})$ the values of the scalar field at the
beginning and the end of inflation are
$(\phi_i,\phi_f)=(1.42088\times 10^{-8},2.41513\times 10^{-16})$,
which are sub-Planckian values, thus the model is self-consistent.
Also note that the singularity in the potential (\ref{potential})
is avoided because it may occur only for $\phi\sim d$, and the
scalar field never reaches the values of $d$ we used, which is
$d\sim 10^{-21}$. More importantly, we find that
$c_T^2-1=1.92608\times 10^{-16}$ and also the model well satisfies
the constraint (\ref{constraintclass}), since we find that for
$(c,d,M,\lambda)=(10^{25.5},10^{-21},1.46725\times
10^{-22},10^{-16.47})$, $c\, \xi H^2\sim \mathcal{O}(10^4)$.
Moreover, let us check whether the slow-roll approximations
(\ref{slowroll}) and also the approximations of Eq.
(\ref{approximationsnatural}) hold true. As we found, for
$(c,d,M,\lambda)=(10^{25.5},10^{-21},1.46725\times
10^{-22},10^{-16.47})$ we have
$\dot{\phi}^2\sim\mathcal{O}(10^{-42})$, $c\, \xi\dot{\phi}^2\sim
\mathcal{O}(10^{-38})$ and $V\sim \mathcal{O}(10^{-22})$, hence
the model is self-consistent and well fitted within the Planck
2018 data. Let us also study the behavior of the gravitational
wave speed as a function of the $e$-foldings number. Before that
however, we need to note that the gravitational wave speed must
satisfy the GW170817 constraints only at first horizon crossing,
since after that and especially near the end of inflation, the
slow-roll conditions are violated so it does not make much sense
to use the expressions for the slow-roll indices and the other
quantities we used in the previous section. However, for
completeness, in Fig. \ref{plotreff} we present the behavior of
the gravitational wave speed $c_T^2$. As it can be seen, it is
equal to unity, but this should not be taken for granted, since
the slow-roll assumptions are violated near the end of inflation
and thus the expressions we used might not be applicable.
Let us consider another viable potential at this point, in which
case the potential is,
\begin{equation}\label{potential1}
V(\phi)=M \left(1-\frac{d}{ \phi }\right)^{1/100}\, ,
\end{equation}
in which case the scalar coupling function $\xi(\phi)$ reads,
\begin{equation}\label{xiphi1}
\xi(\phi)=\frac{d \lambda }{100 M \phi ^2 \left(1-\frac{d}{\phi
}\right)^{101/100}}\, ,
\end{equation}
so the first slow-roll index reads in this case,
\begin{equation}\label{slowrollindexena1}
\epsilon_1=\frac{d}{1600 c\, d \lambda  \phi -1600 c \lambda \phi
^2}\, ,
\end{equation}
and thus by solving the equation $|\epsilon_1(\phi_f)|=1$ we
obtain the value of the scalar field at the end of inflation
$\phi_f$ which is,
\begin{equation}\label{phif1}
\phi_f=\frac{\sqrt{400 c^2 d^2 \lambda ^2-c\, d \lambda }+20 c\, d
\lambda }{40 c \lambda }\, ,
\end{equation}
and from Eq. (\ref{finalinitialefoldings}), the value of the
scalar field at first horizon crossing is,
\begin{equation}\label{phii1}
\phi_i=\frac{20 c\, d \lambda +\sqrt{c\, d \lambda (400 c\, d
\lambda -1)}+10 N}{40 c \lambda }\, .
\end{equation}
Now a viable phenomenology is obtained for $N=61$ for various
values of the free parameters, for example if we choose the free
parameters of the model as follows
$(c,d,M,\lambda)=(10^{25.5},-10^{-21},1.48126\times
10^{-24},10^{-16.47})$, in which case, the spectral index of the
scalar perturbations, the tensor-to-scalar ratio and the tensor
spectral index take the values $n_{\mathcal{S}}=0.967466$,
$r=4.57199 \times 10^{-17}$ and $n_{\mathcal{T}}=-5.74661\times
10^{-18}$, thus the model is viable regarding the observational
indices of inflation. Also in Fig. \ref{plot2} we confront the
model with the Planck likelihood curves, for $N=[50,65]$ and we
can see that the model can be well fitted in the Planck data.
Also the amplitude of the scalar perturbations for the model at
hand for $(c,d,M,\lambda)=(10^{25.5},-10^{-21},1.48126\times
10^{-24},10^{-16.47})$ is $\mathcal{P}_{\zeta}(k_*)=2.196\times
10^{-9}$, and actually, the value of the parameter $M$ crucially
affects only the amplitude of the scalar perturbations, since all
the rest of the inflationary indices are independent of $M$. Now,
for $(c,d,M,\lambda)=(10^{25.5},-10^{-21},1.48126\times
10^{-24},-10^{-16.47})$ the values of the scalar field at the
beginning and the end of inflation are
$(\phi_i,\phi_f)=(1.42088\times 10^{-8},2.41508\times 10^{-17})$,
which are sub-Planckian values, hence the model is
self-consistent. Also note that the singularity in the potential
(\ref{potential1}) is avoided because $d<0$ in this case. Also, we
find that $c_T^2-1=1.92608\times 10^{-18}$ and in addition the
model well satisfies the constraint (\ref{constraintclass}), since
we find that for
$(c,d,M,\lambda)=(10^{25.5},-10^{-21},1.48126\times
10^{-24},10^{-16.47})$, $c\, \xi H^2\sim \mathcal{O}(10^2)$.
Furthermore, let us check whether the slow-roll approximations
(\ref{slowroll}) and in addition the approximations of Eq.
(\ref{approximationsnatural}) hold true for this potential. As we
found, for $(c,d,M,\lambda)=(10^{25.5},-10^{-21},1.48126\times
10^{-24},10^{-16.47})$ we have
$\dot{\phi}^2\sim\mathcal{O}(10^{-42})$, $c\, \xi\dot{\phi}^2\sim
\mathcal{O}(10^{-44})$ and $V\sim \mathcal{O}(10^{-24})$,
therefore the model at hand is self-consistent.

\begin{figure}
\centering
\includegraphics[width=25pc]{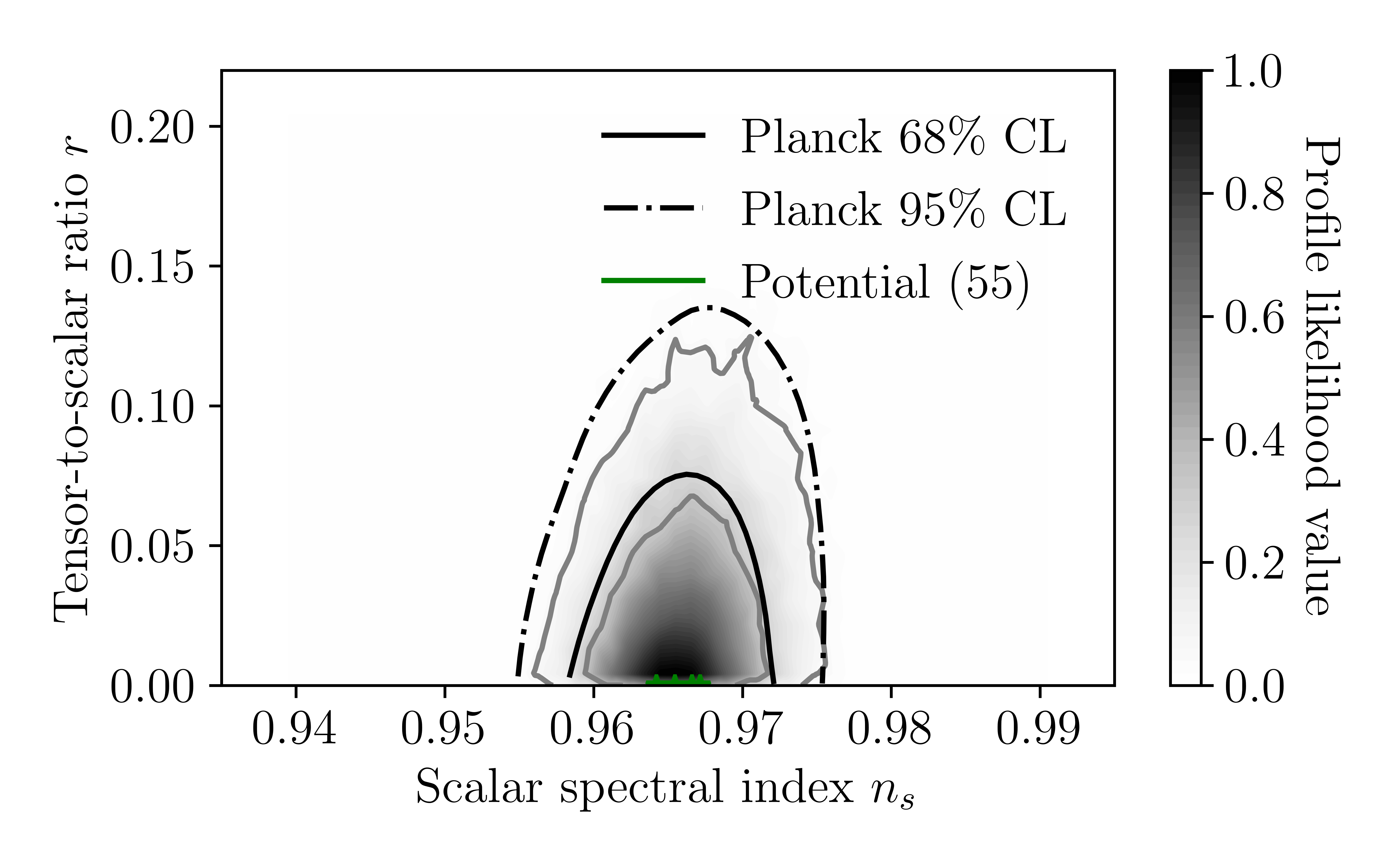}
\caption{Marginalized curves of the Planck 2018 data and the model
with potential $V(\phi)=M \left(1-\frac{d}{ \phi }\right)^2$ of
Eq. (\ref{potential2}) vs the Planck 2018 data (green curve) for
for $N=[50,65]$ and
$(c,d,M,\lambda)=(10^{25.5},-10^{-21},1.48126\times
10^{-24},10^{-16.47})$.}\label{plot3}
\end{figure}

Another viable potential is the following,
\begin{equation}\label{potential2}
V(\phi)=M \left(1-\frac{d}{ \phi }\right)^2\, ,
\end{equation}
in which case the scalar coupling function $\xi(\phi)$ reads,
\begin{equation}\label{xiphi2}
\xi(\phi)=\frac{2 d \lambda }{M \phi ^2 \left(1-\frac{d}{\phi
}\right)^3}\, ,
\end{equation}
so in this case the first slow-roll index reads,
\begin{equation}\label{slowrollindexena2}
\epsilon_1=\frac{d}{8 c\, d \lambda  \phi -8 c \lambda  \phi ^2}\,
,
\end{equation}
and thus in this case by solving the equation
$|\epsilon_1(\phi_f)|=1$ we obtain $\phi_f$ which is,
\begin{equation}\label{phif2}
\phi_f=\frac{\sqrt{2} \sqrt{2 c^2 d^2 \lambda ^2-c\, d \lambda }+2
c\, d \lambda }{4 c \lambda }\, ,
\end{equation}
and from Eq. (\ref{finalinitialefoldings}), $\phi_i$ is,
\begin{equation}\label{phii2}
\phi_i=\frac{2 c\, d \lambda +\sqrt{2} \sqrt{c\, d \lambda  (2 c\,
d \lambda -1)}+N}{4 c \lambda }\, .
\end{equation}
Now a viable phenomenology can be obtained for $N=61$ for various
values of the free parameters in this case too, for example if we
choose the free parameters of the model as follows
$(c,d,M,\lambda)=(10^{25.5},-\frac{1}{10^{21}},2.93436\times
10^{-22}),10^{-17.25}$, in which case, the spectral index of the
scalar perturbations, the tensor-to-scalar ratio and the tensor
spectral index take the values $n_{\mathcal{S}}=0.967494$,
$r=1.51752 \times 10^{-15}$ and $n_{\mathcal{T}}=-1.9074\times
10^{-16}$, thus the model is viable regarding the observational
indices of inflation. Also in this case, the model is viable for
$d<0$ so the singularity in the potential is avoided. Also in Fig.
\ref{plot3} we confront the model with the Planck likelihood
curves, for $N=[50,65]$ and we can see that the model can be well
fitted in the Planck data.
Also the amplitude of the scalar perturbations for the model at
hand for
$(c,d,M,\lambda)=(10^{25.5},-\frac{1}{10^{21}},2.93436\times
10^{-22}),10^{-17.25}$ is $\mathcal{P}_{\zeta}(k_*)=2.196\times
10^{-9}$, and actually, the value of the parameter $M$ crucially
affects only the amplitude of the scalar perturbations, since all
the rest of the inflationary indices are independent of $M$. Now,
for $(c,d,M,\lambda)=(10^{25.5},-\frac{1}{10^{21}},2.93436\times
10^{-22}),10^{-17.25}$ the values of the scalar field at the
beginning and the end of inflation are
$(\phi_i,\phi_f)=(8.56165\times 10^{-8},8.38407\times 10^{-16})$,
which are sub-Planckian values, thus the model is self-consistent.
More importantly, we find that $c_T^2-1=6.39299\times 10^{-17}$
and also the model well satisfies the constraint
(\ref{constraintclass}), since we find that for
$(c,d,M,\lambda)=(10^{25.5},-\frac{1}{10^{21}},2.93436\times
10^{-22}),10^{-17.25}$, $c\, \xi H^2\sim \mathcal{O}(10^2)$.
Moreover, let us check whether the slow-roll approximations
(\ref{slowroll}) and also the approximations of Eq.
(\ref{approximationsnatural}) hold true. As we found, for
$(c,d,M,\lambda)=(10^{25.5},-\frac{1}{10^{21}},2.93436\times
10^{-22}),10^{-17.25}$ we have
$\dot{\phi}^2\sim\mathcal{O}(10^{-40})$, $c\, \xi\dot{\phi}^2\sim
\mathcal{O}(10^{-38})$ and $V\sim \mathcal{O}(10^{-22})$,
therefore the model is self-consistent and well fitted within the
Planck 2018 data.

Let us quote at this point, several other potentials we found that
these are viable, without giving many details for brevity. For
example, a similar phenomenology with the above potentials is
obtained by the potential,
\begin{equation}\label{potential3}
V(\phi)=M \left(1-\frac{d}{\phi }\right)^{100}\, .
\end{equation}
A rather disturbing feature for this class of models is that the
gravitational wave speed is larger than the speed of light, at an
acceptable level, but still, it is a mentionable and disturbing
feature. Another viable potential of this sort with the scalar
coupling function related to the potential as in Eq.
(\ref{couplingfunctionchoices3}) is the following,
\begin{equation}\label{othertwopotentials}
V(\phi)=M \phi^n\, ,
\end{equation}
which however predicts a short inflationary era for viability. For
example for $N=30$ $e$-foldings, and for the free parameters
chosen as
$(c,M,\lambda,n)=(10^{40},-\frac{1}{10^{21}},2.93436\times
10^{-22},10^{16},10^{-15})$, one gets $n_{\mathcal{S}}=0.967638$,
$r=1.28053\times 10^{-16}$ and $c_T^2-1=-5.39374\times 10^{-18}$,
thus in this case the gravitational wave speed is smaller than the
speed of light, and almost equal to unity. Short viable
inflationary eras can be obtained by another class of models, that
of Eq. (\ref{couplingfunctionchoices1}). We examined the potential
$V(\phi)=V_0\phi^n$ and we found viability of the model for
$N=30$, for various values of the free parameters, but we omit the
details for brevity.

\subsection{Phenomenology of Unconstrained Models}

Now let us consider the unconstrained models which as it proves
have less ability to provide viable phenomenologies for a normal
duration of the inflationary era. In this case, the first
slow-roll index takes the form,
\begin{equation}\label{firstslowrolluncostrainedanalytic}
\epsilon_1=\frac{3 V'(\phi )^2}{2 V(\phi ) \left(4 \sqrt{3} c \xi
(\phi ) V(\phi )^{3/2}+\sqrt{3} \sqrt{V(\phi )}\right)^2}-\frac{3
c \xi (\phi ) V'(\phi )^2}{\left(4 \sqrt{3} c \xi (\phi ) V(\phi
)^{3/2}+\sqrt{3} \sqrt{V(\phi )}\right)^2}\, ,
\end{equation}
and the expression entering the $e$-foldings number has the form,
\begin{equation}\label{functionalformefoldings}
\frac{V(\phi ) (4 c \xi (\phi ) V(\phi )+1)}{V'(\phi )}\, ,
\end{equation}
so the functional forms of the kinetic coupling function
$\xi(\phi)$ used in the previous section, greatly simplify the
above. We shall present one viable model which we found for
brevity, using the functional form of $\xi(\phi)$ presented in Eq.
(\ref{couplingfunctionchoices1}). We consider the following
potential at this point,
\begin{equation}\label{potential9}
V(\phi)=\mu  \phi^{-n}\, ,
\end{equation}
and by choosing $\xi(\phi)$ as in Eq.
(\ref{couplingfunctionchoices1}) the scalar coupling function
$\xi(\phi)$ reads,
\begin{equation}\label{xiphi9}
\xi(\phi)=\lambda  \phi ^n\, ,
\end{equation}
so the first slow-roll index reads in this case,
\begin{equation}\label{slowrollindexena9}
\epsilon_1=\frac{n^2 (1-2 c \lambda  \mu )}{2 (4 c \lambda \mu
\phi +\phi )^2}\, ,
\end{equation}
and therefore by solving the equation $|\epsilon_1(\phi_f)|=1$ we
get the value of the scalar field at the end of inflationary era
$\phi_f$ which is,
\begin{equation}\label{phif9}
\phi_f=\frac{\sqrt{n^2-2 c \lambda  \mu  n^2}}{\sqrt{2} \sqrt{16
c^2 \lambda ^2 \mu ^2+8 c \lambda  \mu +1}}\, ,
\end{equation}
and by using Eq. (\ref{finalinitialefoldings}), the value of the
scalar field at first horizon crossing is in this case,
\begin{equation}\label{phii9}
\phi_i=\frac{\sqrt{-2 c \lambda  \mu  n-4 (4 c \lambda \mu
N+N)+n}}{\sqrt{2} \sqrt{\frac{(4 c \lambda \mu +1)^2}{n}}}\, .
\end{equation}
Now a viable phenomenology is obtained in this case for a short
inflationary era with $N=30$ for various values of the free
parameters, for example if we choose the free parameters of the
model at hand as follows
$(c,n,\mu,\lambda)=(10^{30},10^{-15},2.11\times 10^{-25},0.1)$,
the spectral index of the scalar perturbations, the
tensor-to-scalar ratio and the tensor spectral index take the
values $n_{\mathcal{S}}=0.966667$, $r=6.59284 \times 10^{-18}$ and
$n_{\mathcal{T}}=-8.28735\times 10^{-19}$, hence the model is
viable regarding the observational indices of inflation. Also the
amplitude of the scalar perturbations for the model at hand for
$(c,n,\mu,\lambda)=(10^{30},10^{-15},2.11\times 10^{-25},0.1)$ is
$\mathcal{P}_{\zeta}(k_*)=2.19063\times 10^{-9}$, and actually,
the value of the parameter $\mu$ crucially affects only the
amplitude of the scalar perturbations. Now, for
$(c,n,\mu,\lambda)=(10^{30},10^{-15},2.11\times 10^{-25},0.1)$ the
values of the scalar field at the beginning and the end of
inflation are $(\phi_i,\phi_f)=(2.73863\times
10^{-10},1.76781\times 10^{-19})$, which are in this case too
sub-Planckian values, thus the model is self-consistent. More
importantly, we find that $c_T^2-1=2.77781\times 10^{-19}$, in
which case we have the disturbing feature of having a
gravitational wave speed to be larger than the speed of light,
even for a slight extent. Finally, let us check whether the
slow-roll approximations (\ref{slowroll}) and also the
approximations of Eq. (\ref{approximationsnatural}) hold true. As
we found, for $(c,n,\mu,\lambda)=(10^{30},10^{-15},2.11\times
10^{-25},0.1)$ we have $\dot{\phi}^2\sim\mathcal{O}(10^{-48})$,
$c\, \xi\dot{\phi}^2\sim \mathcal{O}(10^{-44})$ and $V\sim
\mathcal{O}(10^{-25})$, hence the model is self-consistent and
also well fitted within the Planck 2018 data.

For completeness, let us mention that many models
$\xi(\phi)=\frac{\lambda  V'(\phi )^2}{V(\phi )}$ yield an
interesting phenomenology, however all the models have a large
tensor-to-scalar ratio, so we will not present these here.

\section{Conclusions}

In this article we thoroughly studied the inflationary
phenomenology of non-minima derivative coupling theories in light
of the GW170817 constraints on the speed of gravitational waves.
Since spacetime is classical during and after the inflationary
era, there is no fundamental reason for the graviton to change its
mass, thus the GW170817 constraints even the primordial tensor
perturbations, so in this work we formalized the inflationary
phenomenology of non-minimal derivative coupling theories, taking
also into account the constraints of the GW170817 event. By using
the slow-roll assumptions solely, we provided analytic relations
for the slow-roll indices and the observational indices of
inflation, including the amplitude of scalar perturbations, which
are severely constrained by the latest Planck data. We introduced
several classes of models, using several convenient relations
between the scalar potential $V(\phi)$ and the scalar coupling
function $\xi(\phi)$, and we presented in detail their
phenomenology. We found several viable models, with interesting
features, that are compatible with both the Planck data and the
GW170817 event. Some of the viable models predict a tiny deviation
of the order $\sim \mathcal{O}(10^{-17}-10^{-19})$ from the speed
of light, and actually the propagation speed is slightly larger
from that of light, so these models are slightly ghost theories to
a very tiny extent. Also some viable models predict a short
inflationary period. We presented only some of the possible models
that can be used, and as we demonstrated, these theories can be
considered as viable candidate theories of inflation, because they
also have a theoretical attribute of containing quantum imprints
in the inflationary Lagrangian. What we did not seek in this work
is whether models exist that may lead to a significantly
blue-tilted tensor spectral index, which may have implications for
future gravitational wave observations. This task will be the
focus of a future work. Also another interesting point is whether
the models we presented with a specific scalar Gauss-Bonnet
coupling $\xi(\phi)$ are valid only for the inflationary era, or
these may be used for other evolutionary eras of our Universe, up
to late times. The answer is that these models are valid only for
the slow-roll era during the inflationary era, since for reheating
and subsequent eras, the Hubble rate is different, so given the
potential, the scalar Gauss-Bonnet coupling function $\xi (\phi)$
might be required to be approximated by a distinct function, see
the discussion in Ref. \cite{Oikonomou:2024jqv}.

\section*{Acknowledgments}

This research has been is funded by the Committee of Science of
the Ministry of Education and Science of the Republic of
Kazakhstan (V.K.O) (Grant No. AP14869238).

\end{document}